\documentclass[12pt,titlepage]{article}
\newcommand{\be}{\begin{equation}}
\newcommand{\ee}{\end{equation}}

\newcommand{\magsim}
{\ \lower2pt\hbox{$\sim $}\mkern-14mu \raise2pt\hbox{$>$}\ }

\newcommand{\yr}{{\rm \, yr}}
\newcommand{\cm}{{\rm \, cm}}
\newcommand{\gm}{{\rm \, g}}

\newcommand{\rad}{{\rm \, rad}}
\def\refnew#1{(\ref{#1})}
\begin{document}

\begin{center}
{\tiny .  }
\vskip 0.5truein
{\LARGE \bf Origin of Chaos in the Prometheus-Pandora System}
\vskip 0.5truein
\baselineskip16pt
{\Large \bf Peter Goldreich} \\
\vskip 0.15truein
{\large \it Institute for Advanced Study}\\
{\large \it Princeton NJ 08540}\\
{\large \it E-mail: pmg@sns.ias.edu}\\
\& \\
{\large \it California Institute of Technology} \\
{\large \it Pasadena CA 91125} \\
{\large \it E-mail: pmg@tapir.caltech.edu}\\
\vskip 0.25truein
{\large and}
\vskip 0.25truein
{\Large \bf Nicole Rappaport} \\
\vskip 0.15truein
{\large \it Jet Propulsion Laboratory} \\
{\large \it California Institute of Technology}  \\
{\large \it Pasadena CA 91109} \\
{\large \it E-mail: Nicole.J.Rappaport@jpl.nasa.gov} \\
\large
\vskip 0.5truein
Submitted to Icarus on June 18, 2003 \\
\vskip 0.5truein
\baselineskip20pt
Number of pages: 23 \\
Number of tables: 3 \\
Number of figures: 7

\end{center}

\pagebreak

\noindent {\bf Proposed Running Head:} Origin of Chaos in the 
Prometheus / Pandora System
\vskip 0.5truein
\noindent {\bf Editorial correspondence to:} \\
\baselineskip14pt
\noindent Dr. Nicole J. Rappaport \\
\noindent Jet Propulsion Laboratory \\
\noindent MS 301-150 \\
\noindent Pasadena CA 91109 \\
\noindent Phone: (818) 354-8211 \\
\noindent Fax: (818) 393-6388 \\
\noindent E-mail: Nicole.J.Rappaport@jpl.nasa.gov
\baselineskip29pt
\pagebreak

\begin{center}
{\bf ABSTRACT}
\end{center}

\begin{minipage}{5.5in}
\noindent
\small

We demonstrate that the chaotic orbits of Prometheus and Pandora are
due to interactions associated with the 121:118 mean motion
resonance. Differential precession splits this resonance into a
quartet of components equally spaced in frequency.  Libration widths
of the individual components exceed the splitting resulting in
resonance overlap which causes the chaos. A single degree of freedom
model captures the essential features of the chaotic dynamics. Mean
motions of Prometheus and Pandora wander chaotically in zones of width
$1.8\deg\yr^{-1}$ and $3.1\deg\yr^{-1}$, respectively.

\vskip 0.25truein

{\bf Key Words:} Satellites of Saturn, Orbits, Chaos

\normalsize
\end{minipage}

\pagebreak

\section{INTRODUCTION}

Goldreich and Rappaport (2003) (hereafter abbreviated as GR) showed
that the motions of Prometheus and Pandora are chaotic.  The chaos
arises from their mutual gravitational interactions, which explains
why their longitude discrepancies have comparable magnitudes and
opposite signs (French {\it et al.} 2002). Numerical integrations that
account for the full mutual interactions and Saturn's gravitational
oblatenss yield a Lyapunov exponent of order $0.3 \yr^{-1}$.  Although
the results reported by GR assumed satellite masses based on a nominal
density of $0.63 \gm\cm^{-3}$, the Lyapunov exponent is insensitive to
the assumed density above a critical value of approximately
$0.3\gm\cm^{-3}$.

GR's integrations also reproduce qualitative features of the
discrepancies between the longitudes of the satellites derived from
analysis of recent HST data and predictions based on orbits fit to
Voyager images (French {\it et al.}  2002). Sudden changes in the mean
motions of Prometheus and Pandora are a striking feature of the
numerical integrations. These occur at intervals of $6.2 \yr$ when the
satellites' apses are anti-aligned. It is notable that the only
clearly documented changes in the mean motions occurred around the
time of the most recent apse anti-alignment ({\it cf.} GR).

The plan of this paper is as follows.  In \S 2 a quartet of 121:118
mean motion resonances is identified as the probable cause of the
chaos.\footnote{Differential apsidal precession splits each mean
motion resonance into a multiplet of closely spaced components.}  Then
we describe two new programs in which interactions between the
satellites are limited to those due to this quartet. The simpler of
these reduces the resonant dynamics to a system with one degree of
freedom. Results from these programs are compared in \S 3 with those
obtained from integrations that account for the full gravitational
interactions. \S 4 is devoted to a discussion of the width of the
chaotic zone.

\section{ORIGIN OF CHAOS}

\subsection{Resonant Quartet}

A systematic search for $j:j-k$ mean motion resonances with $k\leq 4$ 
turned up $j = 121,\;\; k = 3$.

Following Murray and Dermott (2001), we write the disturbing function 
for the action of Pandora on Prometheus as
\begin{equation}
{\mathcal R}  = \frac{Gm'}{a'} R_D\;\;, \label{pandonprom}
\end{equation}
and that for the action of Prometheus on Pandora as
\begin{equation}
{\mathcal R'} = \frac{Gm}{a'} R_D\;\;. \label{promonpand}
\end{equation}
Here $m$ and $a$ denote mass and semi-major axis, and $G$ is the 
gravitational constant.\footnote{Primed and unprimed symbols refer to 
Prometheus and Pandora, respectively.}  To lowest order in the 
eccentricities, the terms in the literal expansion of the disturbing 
function associated with a $k=3$ resonance take the form
\begin{eqnarray}
R_D =
&& e^3 \ f_{82} \ \cos \left[121 \lambda' - 118 \lambda - 3 \varpi
\right] + \nonumber \\
&& e^2 e' \ f_{83} \ \cos \left[ 121 \lambda' - 118 \lambda - 2 \varpi
- \varpi' \right] + \nonumber \\
&& e e'^2 \ f_{84} \ \cos \left[ 121 \lambda' - 118 \lambda - \varpi
- 2 \varpi' \right] + \nonumber \\
&& e e'^3 \ f_{85} \ \cos \left[ 121 \lambda' - 118 \lambda - 3 \varpi'
\right], \label{rd}
\end{eqnarray}
where $e$, $\lambda$, and $\varpi$ stand for eccentricity, mean 
longitude, and argument of periapse.  The $f_{8n}$ are expressed in 
terms of Laplace coefficients evaluated at $\alpha = 
a/a'$.\footnote{In the following, $j$ should be viewed as a shorthand 
for $121$.}
\begin{eqnarray}
f_{82} = && \frac{1}{48} \left\{
\left( -26j + 30 j^2 - 8 j^3 \right) b_{1/2}^{(j)} (\alpha) +
\left( -9 + 27 j - 12 j^2 \right)
\alpha \frac{d b_{1/2}^{(j)} (\alpha)}{d \alpha} + \right. \nonumber \\
&& \left. \left( 6 - 6 j \right)
\alpha^2 \frac{d^2 b_{1/2}^{(j)} (\alpha)}{d \alpha^2} -
\alpha^3 \frac{d^3 b_{1/2}^{(j)} (\alpha)}{d \alpha^3} \right\},
\label{ftwo}
\end{eqnarray}
\begin{eqnarray}
f_{83} = && \frac{1}{16} \left\{
\left( -9 + 31 j - 30 j^2 + 8 j^3 \right) b_{1/2}^{(j-1)} (\alpha) +
\left( 9 - 25 j + 12 j^2 \right)
\alpha \frac{d b_{1/2}^{(j-1)} (\alpha)}{d \alpha} + \right. \nonumber \\
&& \left. \left( -5 + 6 j \right)
\alpha^2 \frac{d^2 b_{1/2}^{(j-1)} (\alpha)}{d \alpha^2} +
\alpha^3 \frac{d^3 b_{1/2}^{(j-1)} (\alpha)}{d \alpha^3} \right\},
\label{fthree}
\end{eqnarray}
\begin{eqnarray}
f_{84} = && \frac{1}{16} \left\{
\left( 8 - 32 j + 30 j^2 - 8 j^3 \right) b_{1/2}^{(j-2)} (\alpha) +
\left(-8 + 23 j - 12 j^2 \right)
\alpha \frac{d b_{1/2}^{(j-2)} (\alpha)}{d \alpha} + \right. \nonumber \\
&& \left. \left( 4 - 6 j \right)
\alpha^2 \frac{d^2 b_{1/2}^{(j-2)} (\alpha)}{d \alpha^2} -
\alpha^3 \frac{d^3 b_{1/2}^{(j-2)} (\alpha)}{d \alpha^3} \right\},
\label{ffour}
\end{eqnarray}
\begin{eqnarray}
f_{85} = && \frac{1}{48} \left\{
\left(-6 + 29 j -  30 j^2 + 8 j^3 \right) b_{1/2}^{(j-3)} (\alpha) +
\left( 6 - 21 j + 12 j^2 \right)
\alpha \frac{d b_{1/2}^{(j-2)} (\alpha)}{d \alpha} + \right. \nonumber \\
&& \left. \left( -3 + 6 j \right)
\alpha^2 \frac{d^2 b_{1/2}^{(j-3)} (\alpha)}{d \alpha^2} +
\alpha^3 \frac{d^3 b_{1/2}^{(j-3)} (\alpha)}{d \alpha^3} \right\}.
\label{ffive}
\end{eqnarray}

Tables 1 and 2 list values for the parameters used in this
paper. Satellite masses are given as fractions of Saturn's mass based
on an assumed density of $0.63\gm\cm^{-3}$. Initial values for mean
longitudes, apsidal angles, mean motions, and eccentricities are based
on orbits fit to Voyager images by Jacobson (2001). Precession rates
are calculated from the Saturnian gravitational field (Campbell and
Anderson 1989).

\underbar{Table 1}. Masses, Initial Mean Longitudes, \& Mean Motions.

\vskip 0.25truein

\begin{tabular}{|l|c|c|c|c|}
\hline Satellite &$m/M$ &Mean Longitude ($^\circ$)& Mean Motion ($^\circ/s$)\\
\hline Prometheus &$5.80\times 10^{-10}$&188.53815& $6.797331 \times 10^{-3}$\\
Pandora & $3.43\times 10^{-10}$&82.14727& $6.629506 \times 10^{-3}$\\ \hline
\end{tabular}

\vskip 0.5truein

\underbar{Table 2}. Eccentricities, Initial Apsidal Angles,
\& Precession Rates.

\vskip 0.25truein

\begin{tabular}{|l|c|c|c|c|}
\hline Satellite &Eccentricity &Apsidal Angle ($^\circ$)& Precession Rate
($^\circ/s$)\\
\hline Prometheus &$2.29\times 10^{-3}$ &212.85385 & $3.1911\times 10^{-5}$\\
Pandora & $4.37\times 10^{-3} $&68.22910 & $3.0082\times 10^{-5}$\\ \hline
\end{tabular}

\vskip 0.5truein

Rates of change of the arguments, corresponding periods, and 
coefficients for the four terms in equation (\ref{rd}) are given in 
Table 3.

\underbar{Table 3}. Resonance Arguments, Rates of Change, Periods, 
and Coefficients.

\vskip 0.25truein

\begin{tabular}{|l|c|c|c|c|}
\hline
Argument & Rate of Change  ($^{\circ} / s$) & Period (year) &
Coefficient \\
\hline
$121 \lambda' - 118 \lambda - 3 \varpi$ &
$-1.058 \times 10^{-5}$ & 1.078 & $e^3f_{82}=-0.001$ \\
$121 \lambda' - 118 \lambda - 2 \varpi - \varpi'$ &
$-0.875 \times 10^{-5}$ & 1.303 & $e^2e'f_{83}=0.006$ \\
$121 \lambda' - 118 \lambda - \varpi - 2\varpi'$ &
$-0.692 \times 10^{-5}$ & 1.648 & $ee'^2f_{84}=-0.01$ \\
$121 \lambda' - 118 \lambda - 3 \varpi'$ &
$-0.509 \times 10^{-5}$ & 2.239 & $e'^3f_{85}=0.007$ \\
\hline
\end{tabular}

\vskip 0.25truein

\subsection{Numerical Integrations}
\label{subsecnumint}

To demonstrate that the quartet of $121:118$ resonances is the cause
of chaos in the Prometheus Pandora system, we develop two new programs
to integrate the satellites' equations of motion.  Interactions
between the satellites are restricted to the four resonant interaction
terms in the Fourier expansion of the disturbing function $R_D$. Each
program evolves propagates the satellites' orbital elements rather
than their cartesian coordinates and velocities as is done by the
``old program'' FSHEP used in GR.

We adopt epicyclic elliptic elements instead of the more standard 
osculating elliptic elements since, unlike the latter, they do not 
require short period terms to describe elliptic orbits around oblate 
planets (cf. Borderies-Rappaport and Longaretti 1994; henceforth, 
referred to as BRL.).  BRL derive a modified version of Gauss' 
equations for the elements $a_e$, $e_e$, $\varpi_e = \omega_e + 
\Omega_e$, and $\lambda_e=\varpi_e + M_e$.\footnote{Hereafter we drop 
the subscript $e$.}  From these, it is a straightforward exercise to 
derive a restricted version of Lagrange's equations that is valid in 
the planar case.  We work with a simplified set appropriate for low 
eccentricity orbits about a modestly oblate planet. The equations read:
\begin{eqnarray}
\frac{d\lambda}{dt} &=& \Omega\;\; , \label{lagrangeone} \\
&& \nonumber \\
\frac{da}{dt} &=& \frac{2}{\kappa a}
\frac{\partial {\mathcal R}}{\partial \lambda} \;\; , \label{lagrangetwo} \\
&& \nonumber \\
\frac{d \varpi}{dt} &=& \Omega - \kappa\;\; , \label{lagrangethree} \\
&& \nonumber \\
\frac{de}{dt} &=& - \frac{1}{\kappa a^2 e}
\frac{\partial {\mathcal R}}{\partial \varpi}\;\; ,
\label{lagrangefour}
\end{eqnarray}
where
\begin{eqnarray}
\Omega^2 &=& \frac{GM}{a^3}\left[1+{3\over 2}\left(R_p\over
   a\right)^2J_2 -{15\over 8}\left(R_p\over a\right)^4J_4+{35\over
     16}\left(R_p\over a\right)^6J_6\;\; ...\right]\;\;, \label{Omega2} \\
\kappa^2 &=& \frac{GM}{a^3}\left[1-{3\over 2}\left(R_p\over
   a\right)^2J_2 +{45\over 8}\left(R_p\over a\right)^4J_4-{175\over
     16}\left(R_p\over a\right)^6J_6\;\; ...\right]\;\; . \label{kappa2}
\end{eqnarray}

\subsection{New Programs}
\label{subsecFSHEPNEW}

In the planar approximation, the Prometheus-Pandora system has four
degrees of freedom and preserves two integrals, total energy and total
angular momentum. Thus each phase space trajectory lies on a six
dimensional hypersurface embedded in the eight dimensional phase
space.

FSHEPRES integrates the four, first-order equations
\refnew{lagrangeone}-\refnew{lagrangefour} for each satellite. Thus it
differs from FSHEP mainly because it limits the interactions between
the satellites to resonant terms.\footnote{We view as an unimportant
difference the use of orbital elements by FSHEPRES and cartesian
positions and velocities by FSHEP.} Other minor differences arise
because FSHEPRES integrates a simplified set of Lagrange's
equations. In particular, the conservation laws are only approximately
satisfied.

FSHEPSIM integrates only the first two equations \refnew{lagrangeone}
and \refnew{lagrangetwo} for each satellite. This drastic
simplification is reasonable because, as a consequence of the rapid
differential precession caused by Saturn's oblateness, interactions
between the satellites produce negligible effects on their apsidal
angles and orbital eccentricities (GR). A further simplification
arises because conservation of energy implies\footnote{In the
following equation we ignore the interaction energy which is only
significant near conjunctions.}
\begin{equation}
\frac{m}{a^2}\frac{da}{dt}=\frac{m^\prime}{{a^\prime}^2}\frac{da^\prime}{dt}.
\label{Econ}
\end{equation}
Thus the resonant dynamics of the Prometheus-Pandora system reduces to
a single degree of freedom system. It proves convenient to define the
variable
\begin{equation}
\psi=121\lambda^\prime-118\lambda\;\;,\label{defpsi}\;\;\label{psi}
\end{equation}
so that $R_D$ is expressed as
\begin{equation}
R_D=\sum_{k=1}^4 C_k\cos(\psi+\delta_k)\;\; ,\label{RD}
\end{equation}
with each ${\dot\delta}_k=\;$constant. The evolution of $\psi$ is
governed by
\begin{eqnarray}
\frac{d^2\psi}{dt^2}&=&3\left[{(121\,\Omega^\prime})^2\frac{m}{M}+\alpha
(118\,\Omega)^2\frac{m^\prime}{M}\right]\sum_{k=1}^4C_k\sin(\psi+\delta_k)\\
&=&3\times {(121\,\Omega^\prime})^2{m\over M}\left[1+\alpha(m^\prime/m)\right]
\sum_{k=1}^4C_k\sin(\psi+\delta_k)\;\;,
\label{d2psidt2}
\end{eqnarray}
where in writing the second form of equation \refnew{d2psidt2}, we
have applied the mean motion resonance relation
$\Omega^\prime/\Omega\approx 118/121$ and emphasized the contribution
from the lighter body, $m^\prime$. Individual mean longitudes follow
from the relations
\begin{eqnarray}
\lambda(t)&=&\frac{-\alpha (m^\prime/m)\psi(t)+118[\lambda(0)+
\dot{\lambda}(0)t]+121\alpha(m^\prime/m)[\lambda^\prime(0)+
{\dot\lambda}^\prime(0)t]}{121[1+\alpha(m^\prime/m)]}\;\;,\label{lampsi}\\
&& \nonumber \\
\lambda^\prime(t)&=&\frac{\psi(t)+121\alpha(m^\prime/m)[\lambda^\prime(0)+
\dot{\lambda}^\prime(0)t] +118[\lambda(0)+
{\dot\lambda}(0)t]}{121[1+\alpha(m^\prime/m)]}\;\;.\label{lamppsi}
\end{eqnarray}
Although we are left with a system described by a single degree of
freedom, the absence of an energy integral still allows for chaos.

\section{COMPARISON OF RESULTS}
\label{comparison}

In this section we compare result obtained using FSHEPRES and FSHEPSIM
with those obtained with FSHEP.  As in GR, all our simulations are
initialized with orbital elements for Prometheus and Pandora taken
from Jacobson's ephemerides at epoch 1981 August 23 04:02:12
UTC. Comparisons among similar calculations done with each of the
three programs are presented in Figures 1-6. As a consequence of
chaos, qualitative similarities are the best that can be
expected. These are apparent in each set of figures. However, there is
a hint that FSHEPSIM yields a slightly smaller Lyapunov exponent than
either FSHEP or FSHEPRES.

The similarity between the 20 year runs of longitude variations
displayed Figures 1 and 2, while consistent with a Lyapunov exponent
of approximately $0.3\yr^{-1}$ as shown in Figure 3, probably also
reflects the fact that at the Voyager epoch the mean motions of
Prometheus and Pandora were close to their respective maximum and
minimum. This accounts for the negative values of the rates of each
resonant argument quoted in Table 3.

Figure 4 shows that over 3000 years the net variation of $121
\lambda^\prime-118 \lambda-2\varpi^\prime-\varpi$ is much smaller than
that of the other phases.  Together with the constraint imposed by the
conservation of energy on relative variations of $n$ and $n^\prime$,
this implies that over this time interval the average mean value of
$n$ was smaller than its initial value by about $0.67 \deg\yr^{-1}$
and that of $n^\prime$ was larger by about $1.14 \deg\yr^{-1}$. The
relatively small value of $121n^\prime -118
n-2{\dot\varpi}^\prime-{\dot\varpi}$ has a plausible dynamical
explanation. The term with this phase rate is the one with the largest
amplitude. Moreover, the amplitudes of terms with neighboring phase
rates are about half as large and have opposite signs to that of the
dominant term, whereas the amplitude of the term with the slowest
phase rate is much smaller. Support for this explanation is provided
by observing that interchanging the values of $e$ and $e^\prime$
results in the phase $121 \lambda^\prime-118
\lambda-\varpi^\prime-2\varpi$ assuming the special status of being
the one with the smallest net variation.

Figures 5 and 6 display longitude variations over 3000 years relative
to the longitude based on the average mean motion over this interval.
These are seen to be bounded by $\pm 180\;$degrees.

\section{Discussion}

A closer examination of the one-degree of
freedom model developed for FSHEP provides additional insight
regarding chaos in the Prometheus-Pandora system.

Overlapping resonances are known to produce chaos. Frequencies of
individual members of the resonant quartet are spaced by
$\dot\varpi-\dot{\varpi}^\prime\approx 1.0\;\rad\yr^{-1}$. This is
smaller than the half widths of the individual resonance
components.\footnote{The half width is the maximum angular velocity
achieved during motion on the separatrix.}  Half widths computed from
equation \refnew{d2psidt2} and the data in Tables 1-3 are, in order of
increasing resonance frequency, $1.5,\; 3.7,\;
5.1,\;4.1\;\rad\yr^{-1}$.

Figure 7 shows surfaces of section based on data from 3000 year
integrations using FSHEPSIM. A single point with coordinates $\psi -
\varpi - 2 \varpi^{\prime},\; {\dot\psi}-{\dot\varpi}-
2{\dot\varpi}^\prime$ is plotted each time the apses align (every
$6.2\yr$ when $\varpi-\varpi^\prime=0$ modulo $2\pi$).\footnote{We
chose apse alignment to minimize the effects of the interaction
energy.} Nominal values for the satellites' masses were used for
the upper panel. The scattering of points over an area in the phase
plane is a signature of chaos. The balance in the number of
points above and below the horizontal axis and the overall vertical
width of their distribution are a consequence of the dominance of the
resonance component with phase
$121\lambda^\prime-118\lambda-\varpi-2\varpi^\prime$. Satellite
masses were reduced by a factor 10 below their nominal values to
obtain the integrable example whose surface of section is shown in
the lower panel.

Variations of $n$ and $n^\prime$ are related to those of $\dot\psi$ by
\begin{eqnarray}
{dn\over dt}&=&{1\over 121[1+\alpha(m^\prime/m)]}{d\psi\over dt}\;\;,
\label{dotn}\\
&&\nonumber\\
{dn^\prime\over dt} &=&{-\alpha(m^\prime/m)\over 118[1+\alpha
(m^\prime/m)]}{d\psi\over dt}
\;\; .\label{dotnp}
\end{eqnarray}
Thus the total width in $\dot\psi$ corresponds to full width
variations $\Delta n\approx 1.8\deg\yr^{-1}$ and $\Delta
n^\prime\approx 3.1\deg\yr^{-1}$.

\section{Acknowledgments}

This research was supported by NASA Planetary Geology and Geophysics
grant 344-30-53-02 and by NSF grant AST-0098301.

\vfil\eject

\section{References}

\baselineskip16pt

\hskip 0.5truein Borderies-Rappaport N., and P.-Y. Longaretti 1994:
Test Particle Motion around an Oblate Planet, {\it Icarus} {\bf 107},
129-141.

\vskip 0.25truein

\hskip 0.25truein Campbell, J.K., and J.D. Anderson 1989: Gravity
Field of the Saturnian System from {\it Pioneer} and {\it Voyager}
Tracking Data, {\it Astron. J.} {\bf 97}, 1485-1495.

\vskip 0.25 truein

\hskip 0.25truein French, R.G., C.A. McGhee, L. Dones, and J.J.
Lissauer 2003: Saturn's wayward shepherds: the perigrinations of
Prometheus and Pandora, {\it Icarus} {\bf 162}, 144-171.

\vskip 0.25truein

\hskip 0.25truein Goldreich, P. and N. Rappaport 2003: Chaotic
Motions of Prometheus and Pandora, {\it Icarus} {\bf 162}, 391-399.

\vskip 0.25truein

\hskip 0.25truein Jacobson 2001: Tables of Prometheus and Pandora
Planetocentric Mean Elements at Julian Ephemeris Date 2444839.6682
Referred to the Earth Mean Equator and Equinox of J2000 System, {\it
Personal Communication}.

\vskip 0.25truein

\hskip 0.25truein Murray, C.D., and S.F. Dermott 2001:
\underline{Solar System Dynamics}, Cambridge University Press, 592 pp.

\vfil\eject

\section{Figure Captions}

\vskip0.5truecm

\noindent\underbar{FIGURE 1}: Prometheus longitude in degrees
from numerical integration as a function of time over 20 years. A
drift based on the initial mean motion has been subtracted from the
longitudes.  Dashed lines indicate the times of periapsis
antialignment. Results shown in the top, middle, and bottom panels
were obtained with the programs FSHEP, FSHEPRES, and FSHEPSIM.

\vskip0.5truecm

\noindent\underbar{FIGURE 2}: Pandora longitude in degrees from
numerical integration as a function of time over 20 years. A drift
based on the initial mean motion has been subtracted from the
longitudes.  Dashed lines indicate the times of periapsis
antialignment. Results shown in the top, middle, and bottom panels
were obtained with the programs FSHEP, FSHEPRES, and FSHEPSIM.

\vskip0.5truecm

\noindent\underbar{FIGURE 3}: Lyapunov exponent in $\yr^{-1}$ for the
Prometheus-Pandora system over a period of $3\times 10^3$ years. The
results shown in the top, middle, and bottom panels were obtained with
the programs FSHEP, FSHEPRES, and FSHEPSIM.

\vskip0.5truecm

\noindent\underbar{FIGURE 4}: Phases $\psi+\delta_k, \; k = 1,
\ldots, 4$, in radians along the solution. Results shown in the
top, middle, and bottom panels were obtained with the programs FSHEP,
FSHEPRES, and FSHEPSIM.

\vskip0.5truecm

\noindent\underbar{FIGURE 5}: Prometheus longitude in degrees from
numerical integration as a function of time over 3000 years. A drift
based on the mean motion averaged over 3000 years has been subtracted
from the longitude.  Results shown in the top, middle, and bottom
panels were obtained with the programs FSHEP, FSHEPRES, and FSHEPSIM.

\vskip0.5truecm

\noindent\underbar{FIGURE 6}: Pandora longitude in degrees from
numerical integration as a function of time over a 3000 years. A drift
based on the mean motion averaged over 3000 years has been subtracted
from the longitude.  Results shown in the top, middle, and bottom
panels were obtained with the programs FSHEP, FSHEPRES, and FSHEPSIM.

\vskip0.5truecm

\noindent\underbar{FIGURE 7}: Surfaces of section obtained by plotting
$(\psi - \varpi - 2 \varpi^{\prime}$, $\dot{\psi}-
\dot{\varpi}-2\dot{\varpi'})$ at each time of periapsis alignment over
3,000 years. Units are radians and radians per year. Computations were
made with FSHEPSIM, for the top panel with the nominal value of
$0.63\gm\cm^{-3}$ for the satellite density. For the bottom panel, the
density was reduced by a factor of 10 in order to obtain an integrable
example to contrast with the chaotic one shown above.

\end{document}